\documentclass[12pt]{article}

\usepackage{amsmath,amssymb}
\usepackage{graphicx}
\usepackage[pdftex]{hyperref}
\usepackage{epstopdf}
\DeclareGraphicsExtensions{.eps,.bmp,.wmf,.jpeg,.pdf}
\numberwithin{equation}{section}
\def\be{\begin{equation}}
\def\ee{\end{equation}}

\def\bea{\begin{eqnarray}}
\def\eea{\end{eqnarray}}
\title{Power-law expansion and Higgs-type potential in a scalar-tensor model}
\author{L. N. Granda\thanks{ngranda@univalle.edu.co}\\ {\small\it Departamento de Fisica, Universidad del Valle}\\{\small\it A.A. 25360, Cali, Colombia}}
\date{}
\begin{document}
\maketitle

\begin{abstract}
\noindent  In the scalar-tensor model with Gauss-Bonnet and kinetic couplings, the power-law dark energy solution may be described by Higgs-type potential. It was found that in the solution describing early time epoch of matter dominance, the potential presents symmetry breaking phase, and the power law solution leading to accelerated expansion corresponds to Higgs-type potential in its symmetric shape.\\ 

\end{abstract}

\section{Introduction}
\noindent 
The current accelerated expansion of the universe (\cite{perlmutter,riess, kowalski,komatsu,percival,planck1,planck2}) requires among others, a revision of the large scale behavior of gravity. The problem of dark energy (for review see \cite{copeland,sergeiod}) raises fundamental questions like the origin of acceleration and the coincidence problem. In recent years these problems have been addressed in different models, ranging from modifications of the energy-momentum tensor, introducing scalar fields of different nature (see \cite{copeland} and references therein for review), to modifications of the Einstein-Hilbert Lagrangian ($f(R)$ models) \cite{capozziello}, \cite{sodintsov}. 
The scalar-tensor theories  \cite{polarski}, \cite{elizalde}, \cite{peri}, \cite{maeda} are some of the most studied as alternative to gravitation theory and to explain the dark energy. The coupling of scalar field to curvature appears naturally after compatification of higher dimensional theories of gravity such as Kaluza-Klein and string theory, offering the possibility of connecting fundamental scalar fields with the nature of DE. This relationship with the fundamental theories could reveal itself in the current low-curvature universe (see \cite{sergei11} for review). Some late time cosmological aspects of scalar-tensor model with kinetic coupling to curvature have been studied in \cite{sushkov,granda,granda1,gao}.\\
In the present work we study power-law solutions for the scalar-tensor model with kinetic and Gauss Bonnet (GB) couplings  \cite{granda5} with Higgs-type potential. The GB term affects the cosmological dynamics when it is coupled to a dynamically evolving scalar field through arbitrary function of the field, giving rise to second order differential equations of motion (this preserves the theory ghost free) \cite{zwiebach}, \cite{nojiri1}. For more general scalar-tensor theories having second order field equations see \cite{deffayet}, \cite{kobayashi}. The inclusion of coupled GB term enrich the cosmological dynamics of previous models with non-minimal kinetic coupling to curvature \cite{granda}, \cite{granda1}. \\
Of great importance is the the existence of exact power law solutions allowing to explain different phases of the cosmic evolution, when the energy density is modeled by by a perfect
fluid. In the FRW background the power-law solutions represent asymptotic or intermediate states among all possible cosmological evolutions.
We study the implications for the scalar field potential of the considered model, if power-law solutions are assumed to exist. We have found that for the existence of this solution  the potential should be of the Higgs type. This scalar field might be identified with the dark energy field, responsible for
the recent stage of accelerated expansion of the universe. On the other hand, this single scalar field could be coupled with the standard model Higgs scalar, giving rise to a mixing between the Higgs boson and the single scalar, and to the possibility that the standard model Higgs boson (doublet) decay into a pair of singlets \cite{bento}. Nevertheless we ignore this coupling, as in any case it would be very tiny (is out of the current experimental possibilities, since it leaves the Higgs sector of the SM practically unaffected) and we are interested in the dark sector only.\\
This paper is organized as follows. In section II we present the model and the equations of motion in the FRW metric. In section III we derive the expression for the potential compatible with the power-law expansion. In section IV we make an analysis of cosmological perturbations. In section V we present some discussion.
\section{Field Equations}
We consider the following action which adds the Gauss Bonnet coupling to the model with kinetic coupling to curvature considered in \cite{granda}
\be\label{eq1} 
\begin{aligned} 
S=&\int d^{4}x\sqrt{-g}\Big[\frac{1}{16\pi G} R -\frac{1}{2}\partial_{\mu}\phi\partial^{\mu}\phi+\frac{\zeta}{\phi^2}G_{\mu\nu}\partial^{\mu}\phi\partial^{\nu}\phi- V(\phi)-\xi(\phi){\cal G}\Big]
\end{aligned}
\ee
\noindent where $G_{\mu\nu}=R_{\mu\nu}-\frac{1}{2}g_{\mu\nu}R$, ${\cal G}$ is the 4-dimensional GB invariant ${\cal G}=R^2-4R_{\mu\nu}R^{\mu\nu}+R_{\mu\nu\rho\sigma}R^{\mu\nu\rho\sigma}$ . 
One important feature of this model is the dimensionless of the coupling constant $\zeta$, due to the choice of the coupling function as $1/\phi^2$. Any other coupling function would make the coupling constant $\zeta$ dependent on some mass scale which would affect the infrared or ultraviolet behavior of the model when quantum corrections are considered. The properties of the GB invariant guarantee the absence of ghost terms in the theory. Hence, the equations derived from this action contain only second derivatives of the metric and the scalar field. Note also that the kinetic terms in the action can be written in the from $-\frac{1}{2}(g_{\mu\nu}-2\frac{\zeta}{\phi^2}G_{\mu\nu})\partial^{\mu}\phi\partial^{\nu}\phi=-\frac{1}{2}\omega_{\mu\nu}(\phi)\partial^{\mu}\phi\partial^{\nu}\phi$, making possible (in principle) in a given background metric, to redefine the scalar field in such a way to recover the canonical form of the kinetic term (see \cite{sergeio1} where this approach is considered). This will be illustrated in the specific power-law solution considered bellow in section 3.\\
In the spatially-flat Friedmann-Robertson-Walker (FRW) metric
\be\label{eq1a}
ds^2=-dt^2+a(t)^2\left(dr^2+r^2d\Omega^2\right),
\ee
the field equations for this model take the form (see \cite{granda5} for details)
\be\label{eq2}
H^2=\frac{\kappa^2}{3}\rho_{DE}=\frac{\kappa^2}{3}\left(\frac{1}{2}\dot{\phi}^2+V(\phi)+9 \zeta H^2\frac{\dot{\phi}^2}{\phi^2}+24H^3\frac{d\xi}{dt}\right),
\ee
\be\label{eq4}
\begin{aligned}
&\ddot{\phi}+3H\dot{\phi}+\frac{dV}{d\phi}+6\zeta H^2\left(\frac{\ddot{\phi}}{\phi^2}-\frac{\dot{\phi}^2}{\phi^3}\right)
+18 \zeta H^3\frac{\dot{\phi}}{\phi^2}+\\
&12\zeta H\dot{H}\frac{\dot{\phi}}{\phi^2}+24\left(\dot{H}H^2+H^4\right)\frac{d\xi}{d\phi}=0
\end{aligned}
\ee
Here we neglected the matter term since we are considering the limit of scalar field dominance. Due to the kinetic coupling with curvature and the GB coupling, the energy density and pressure derived from the present model will be considered as effective ones. \\
We will assume the existence of exact power-law solution and find the scalar field potential that allows this solution. As will be shown for the model (\ref{eq1}), the power-law solution exists only in the case when we adopt the following Higgs-type form of the potential
\be\label{eq14}
V(\phi)=\frac{1}{2}\mu^2\phi^2+\frac{\lambda}{4}\phi^2
\ee
where $\mu^2<0$ in the solution corresponding to the epoch of matter dominance and $\mu^2>0$ in the phase of accelerated expansion.
\section{The Higgs-type potential and power-law solution}
\noindent {\bf Quintessence power-law}\\
In fact in \cite{granda5} the power-law solution was obtained, and as a result the potential was found of the Higgs-type form (see eq. (3.3) in \cite{granda5} in the case $C=0$). Our main purpose is to highlight this result and additionally to consider power-law solution of the phantom type, and also to relate the symmetric and broken phases of the Higgs-type potential with the character of the power-law expansion (whether it describes presureless mater or accelerated expansion).\\
For the quintessence behavior let's assume the following power-law dependence
\be\label{eq15}
H=\frac{p}{t}, \,\,\,\, \phi(t)=\phi_0\left(\frac{t}{t_1}\right)^{\alpha},\,\,\,\, \xi(t)=\xi_0\left(\frac{t}{t_1}\right)^{\beta}+\xi_1 \ln \frac{t}{t_1}
\ee
where $p>0$ and $\phi_0$, $t_1$, $\xi_0$ $\xi_1$ are constants to be determined by adequate conditions. This solution leads to a constant effective EoS given by
\be\label{eq15a}
w=-1+\frac{2}{3p}
\ee
Replacing (\ref{eq15}) in equations (\ref{eq2}) and (\ref{eq4}), we find the following requirements for the potential 
\be\label{eq16}
\frac{3}{\kappa^2}\frac{p^2}{t^2}=\frac{1}{2}\frac{\alpha^2\phi_0^2}{t_1^{2\alpha}}t^{2\alpha-2}+V+\frac{9\zeta p^2\alpha^2}{t^4}+\frac{24\xi_0\beta p^3}{t_1^{\beta}}t^{\beta-4}+24p^3\xi_1 t^{-4}
\ee
which come from the Eq. (\ref{eq2}), and integrating the equation (\ref{eq4}) with respect to the potential $V$ one finds
\be\label{eq17}
\begin{aligned}
V=&-\frac{(\alpha-1+3p)\alpha^2}{2\alpha-2}\left(\frac{\phi_0}{t_1^{\alpha}}\right)^2t^{2\alpha-2}+\frac{9}{2}\zeta p^2(p-1)\alpha^2 t^{-4}\\
&-\frac{24p^3(p-1)\xi_0\beta}{(\beta-4)t_1^{\beta}}t^{\beta-4}+6p^3(p-1)\xi_1 t^{-4}
\end{aligned}
\ee
where we choose the integration constant equal to zero. In order for the Eqs. (\ref{eq16}) and (\ref{eq17}) to be compatible, the only possible non-trivial values of $\alpha$ and $\beta$ should be 
\be\label{eq17a}
\alpha=-1,\,\,\,\,  \text{and} \,\,\,\,\,\, \beta=2, 
\ee
which lead to the following restrictions
\be\label{eq18}
8\xi_0p(p+1)\kappa^2=t_1^2,\,\,\,\, 6\zeta p(p+1)=-\phi_0^2t_1^2-8p^2(p+3)\xi_1
\ee
Note that from (\ref{eq18}) follows that an appropriate constraint on $\xi_1$ ensures that the sum of kinetic terms in (\ref{eq2}) remain positive. Finally the potential in terms of the scalar field takes the form
\be\label{eq19}
V=\frac{3p(p-1)}{8\xi_0(p+1)^2\kappa^4\phi_0^2}\phi^2+\left(\frac{2p-1}{16\xi_0p(p+1)^2\kappa^2\phi_0^2}-\frac{3\xi_1 p(p-1)}{16\xi_0^2(p+1)^3\kappa^4\phi_0^4}\right)\phi^4
\ee
where we have used (\ref{eq18}). The properties of this potential are correlated with the expansion behavior of the universe. In order to have a Higgs like potential with the phase of broken symmetry, the coefficient of the $\phi^4$ term should be positive, which can be translated into a constraint for the power $p$. We may assume for instance, that the product $\phi_0 \kappa=\phi_0/M_p$ takes the value $\phi_0/M_p\sim 1$ and the constant $\xi_1$ takes the value $\xi_1\sim -\xi_0 (2p-1)/p$. In this case the coefficient of $\phi^4$ remains positive provided that $p>1/2$. The coefficient of $\phi^2$ becomes negative in the interval $1/2<p<1$ and turns positive for $p>1$ . This means that the potential is in the broken symmetry phase in the interval $1/2<p<1$, which includes the epoch of matter dominance where $p=2/3$. The value $p=1$ which corresponds to the divide between the decelerated and accelerated expansion, gives a symmetric potential $V\propto \phi^4$. The values of $p>1$ that lead to accelerated expansion correspond to a symmetric potential. This means that for the present model, in the power-law scenario the DE is related with the Higgs-type potential with unbroken symmetry. According to (\ref{eq15}) and (\ref{eq17a}) the scalar field evolves as $\phi=\phi_0 t_1/t$,
which means that as the universe expands, the scalar field rolls down the potential, and the potential reaches the minimum value (in the case of accelerated expansion) $V=0$ when $\phi=0$, at the limit $t\rightarrow\infty$. An interesting limit of the potential (\ref{eq19}) takes place when $p\rightarrow\infty$, which according to (\ref{eq15a}) is the de Sitter limit of the solution. In this limit the potential becomes quadratic in the scalar field ($V\propto \phi^2$) as the second term in (\ref{eq19}) disappears (assuming that $\xi_1\sim -\xi_0 (2p-1)/p$). \\
This shows that in the frame of the present model, a scalar singlet with a Higgs-type potential gives rise to evolutionary scenarios of the power-law type, where the broken-symmetry potential accounts for the epoch of matter dominance, and the potential in the symmetric configuration gives the appropriate amount of negative pressure to account for the DE. This single scalar field could be coupled with the standard model Higgs scalar, giving rise to a mixing between the Higgs boson and the single scalar, and to the possibility that the standard model Higgs boson (doublet) decay into a pair of singlets \cite{bento}. Nevertheless this coupling would be very tiny (out of the current experimental possibilities) and we are interested in the dark sector only.\\ 
A viable scalar field scenario for DE requires an ultra-light scalar ($m_{\phi}\sim 10^{-33}$ ev.) with an amplitude of the order of the Planck mass ($\phi\sim M_p$) \cite{copeland}. The mass of the scalar field from (\ref{eq19}) for $p>1$ is 
\be\label{eq20a}
m_{\phi}^2=\frac{3p(p-1)}{4(p+1)^2\xi_0\kappa^4\phi_0^2}\sim \frac{M_p^2}{\xi_0}
\ee
where we used $\kappa^2=M_p^{-2}$, and assumed that the field value at present is $\phi_0\sim M_p$ (if $\phi_0$ is the current value, then from (\ref{eq15}) for $\phi$ follows that $t_1$ is of the order of the age of the universe, i.e. $t_1\sim H_0^{-1}$). In order to achieve the required mass for the scalar field, the constant $\xi_0$ should be of the order of $\xi_0\sim M_p^2 H_0^{-2}\sim 10^{120}$. This gives $m_{\phi}\sim H_0\sim 10^{-33} ev$.  Under the same approximations and using $\xi_1=(2p-1)\xi_0/p$ it can be seen from (\ref{eq18}) that $\zeta$ and $\xi_0$ are of the same order. Note that in these approximations the current GB contribution to the total density is of the order of $\rho_{GB}=24\kappa^2H^3(d\xi/dt)\sim \xi_0 H_0^4/M_p^2\sim M_p^2H_0^2$, and the same order of magnitude is valid for the other contributions (up to coefficients that depend on $p$) to the total density that appear in (\ref{eq2}). So the model could be cosmologically viable.\\
\noindent {\bf Phantom power-law}\\
The observational evidence also supports an EoS for the DE with a current value below the cosmological constant divide. Here we consider the following phantom power-law solution
\be\label{eq21}
H=\frac{p}{t_s-t},\,\,\,\, \phi=\phi_0\left(\frac{t_s-t}{t_1}\right)^{\alpha},\,\,\,\,\, \xi=\xi_0\left(\frac{t_s-t}{t_1}\right)^{\beta}+\xi_1 \ln\left(\frac{t_s-t}{t_1}\right)
\ee
which gives the effective EoS parameter
\be\label{eq21a}
w=-1-\frac{2}{3p}
\ee
It is well known that this solution presents future Big Rip singularity at $t=t_s$. Proceeding in the same manner as in the previous case, by replacing in Eqs. (\ref{eq2}) and (\ref{eq4}), we find that the only consistent values for $\alpha$ and $\beta$ are $\alpha=-1$, $\beta=2$  which lead to the following restrictions in order to consistently solve the Eqs. (\ref{eq2}) and (\ref{eq4}) with respect to the potential
\be\label{eq22}
8\xi_0p(p-1)\kappa^2=t_1^2,\,\,\,\, 6\zeta p(p-1)=-\phi_0^2t_1^2+8p^2(p-3)\xi_1
\ee
giving the potential
\be\label{eq23}
V=\frac{3p(p+1)}{8\xi_0(p-1)^2\kappa^4\phi_0^2}\phi^2+\left(\frac{2p+1}{16\xi_0p(p-1)^2\kappa^2\phi_0^2}+\frac{3p(p+1)\xi_1}{16\xi_0^2(p-1)^3\kappa^4\phi_0^4}\right)\phi^4
\ee
The potential is only in the symmetric configuration as the coefficients of $\phi^2$ and $\phi^4$ are always positive (the constants $\xi_0$ and $\xi_1$ should be positive). In this case, values of $0<p<1$ are not allowed according to first of Eqs. (\ref{eq22}), and the kinetic coupling constant $\zeta$ from the second of Eqs. (\ref{eq22}) is positive for values of $p>3$ (in fact, according to observations an appropriate value for $p$ could be $p\sim 27$, which gives $w\sim -1.025$),  which means that in the model (\ref{eq1}) we can have phantom expansion preserving the positive sign of the kinetic term, i.e. without resorting to a phantom scalar field (see Eq. (\ref{eq23b}) below). It then follows that the effective EoS takes values in the interval $-5/3<w<-1$ for $1<p<\infty$. The de Sitter limit takes place for $p\rightarrow\infty$, where the second term in (\ref{eq23}) disappears and the potential becomes $V\propto\phi^2$. \\
The above results show that the Higgs-type potential may be used for description of matter dominated universe (decelerated expansion) if the potential is in the configuration with broken symmetry, and describes dark energy if the potential becomes symmetric provided $p>1$ in the quintessence or phantom phases. In Fig. 1 we outline the shape of the potential according to the value of the power $p$, for all the cases considered above.
\begin{center}
\includegraphics [scale=0.7]{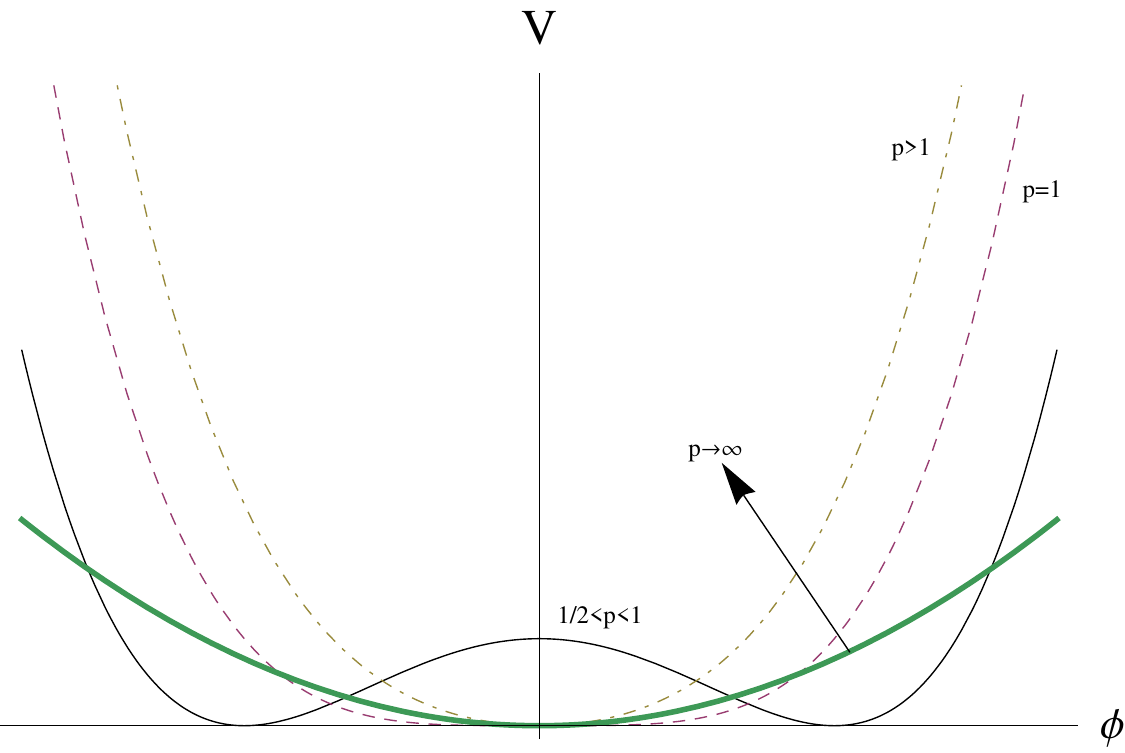}
\end{center} 
\begin{center}
{Fig. 1 \it The shape of the potential according to the different power-law scenarios discussed above.}
\end{center}
Is worth noting that the kinetic terms in the action (\ref{eq1}) can be written in the form
\be\label{eq23a}
-\frac{1}{2}(g_{\mu\nu}-2\frac{\zeta}{\phi^2}G_{\mu\nu})\partial^{\mu}\phi\partial^{\nu}\phi=-\frac{1}{2}\omega_{\mu\nu}(\phi)\partial^{\mu}\phi\partial^{\nu}\phi
\ee
which in the FRW metric gives
\be\label{eq23b}
\frac{1}{2}\omega(\phi)\dot{\phi}^2=\frac{1}{2}\left(1+\frac{18\zeta H^2}{\phi^2}\right)\dot{\phi}^2
\ee
and consequently by redefining the scalar field as \cite{sergeio1}
\be\label{eq23c}
\varphi=\int \sqrt{\omega(\phi)}d\phi
\ee
one may write the kinetic term in the canonical form. For the specific case of the power-law solutions (\ref{eq15}) and (\ref{eq21}) one may note that $\omega=1+\frac{18\zeta p^2}{\phi_0^2 t_1^2}=const.$, and therefore according to (\ref{eq23c}), the scalar field transforms proportional to itself and the shape of the potential remains the same after this transformation. Note that to maintain the correct sign of the kinetic term and to avoid phantom scalar $\omega(\phi)$ should be positive in (\ref{eq23b}) \cite{germani1}, \cite{germani2}. According to second of Eqs. (\ref{eq18}), the constant $\xi_1<0$ is necessary to guarantee that $\omega>0$ for the power-law solution with EoS $w>-1$ (if $\xi_1=0$ in (\ref{eq18}), then from (\ref{eq23b}) follows that $\omega=(1-2p)/(p+1)$, which is negative for $p>1/2$ and leads to ghosts). For the case of phantom power-law, from second of restrictions (\ref{eq22}) we find the same situation except that the coupling constant $\xi_1$ should be positive so that $\omega>0$.
\section{Relation with Galileon theories and metric perturbations}
It is interesting to establish a relationship between the present model and more general models known as Galileon theories. In fact the present model is a particular case of the generalized Galileon theory \cite{deffayet}, \cite{kobayashi}, \cite{felice}, \cite{felice1}, \cite{felice2}. The action for the generalized Galileon model may be written as \cite{kobayashi} 
\be\label{eq24}
\begin{aligned}
&S=\int d^4x\sqrt{-g}\Big[K(\phi,X)-G_3(\phi,X)\Box\phi+G_4(\phi,X)R+\frac{\partial G_4}{\partial X}\left((\Box\phi)^2-(\nabla_{\mu}\nabla_{\nu}\phi)^2 \right)\\
&+G_5(\phi,X)G_{\mu\nu}\nabla_{\mu}\nabla^{\nu}\phi-\frac{1}{6}\frac{\partial G_5}{\partial X}\left[(\Box\phi)^3-3(\Box\phi)(\nabla_{\mu}\nabla_{\nu}\phi)^2+2(\nabla_{\mu}\nabla_{\nu}\phi)^3\right]\Big]
\end{aligned}
\ee
where $X=-\nabla_{\mu}\phi\nabla^{\mu}\phi/2$ and $(\nabla_{\mu}\nabla_{\nu}\phi)^3=(\nabla_{\mu}\nabla_{\nu}\phi)(\nabla^{\nu}\nabla^{\sigma}\phi)(\nabla_{\sigma}\nabla^{\mu}\phi)$. The model (\ref{eq1}) is obtained from (\ref{eq24}) by using the following correspondence \cite{kobayashi}. 
\be\label{eq25}
\begin{aligned}
&K(\phi,X)=-V(\phi)+X-8\frac{d^4\xi}{d\phi^4} X^2(3-\ln X),\,\,\,\, G_3=-4\frac{d^3\xi}{d\phi^3}X(7-3\ln X)\\
&G_4=\frac{1}{2\kappa^2}-4\frac{d^2\xi}{d\phi^2}X(2-\ln X),\,\,\,\, G_5=\frac{\zeta}{\phi}+4\frac{d\xi}{d\phi}\ln X
\end{aligned}
\ee
where the expression for $G_5$ gives (up to total derivative) the non-minimal kinetic coupling as appears in the third term in (\ref{eq1}), and all the terms depending on $\xi$ in (\ref{eq25}) reproduce the GB coupling in (\ref{eq1}). The Galileon and generalized Galileon theories as is also the case of the present model, have the attractive feature that field equations contain derivatives only up to second order, which guarantees the absence of ghosts. However, in a curved space-time background like the FRW, the absence of such instabilities is not guaranteed.\\ 
In view of the importance of the power-law solutions in the DE problem, it is worthwhile to study the stability properties of the solution in our case. Here we will use the perturbative approach developed in \cite{kobayashi}, \cite{felice}, \cite{felice1}, \cite{felice2} for generalized Galileon theories, in order to study the conditions to avoid ghost and gradient instabilities under scalar and tensor perturbations of the metric. In the present work we will illustrate the method for the quintessence power-law solutions.\\
\noindent {\bf \it Tensor perturbations}\\
The general quadratic action for the tensor perturbations $h_{i,j}$ is given by (see \cite{kobayashi})
\be\label{eq26}
\delta^2S_T=\frac{1}{8}\int dt d^3x a^3\left({\cal G}_T\dot{h}_{ij}^2-\frac{{\cal F}_T}{a^2}(\vec{\nabla} h_{ij})^2\right)
\ee
where ${\cal G}_T$ and ${\cal F}_T$ are given by \cite{kobayashi}
\be\label{eq27}
\begin{aligned}
&{\cal G}_T=2\left[G_4-2X\frac{\partial G_4}{\partial X}-X\left(H\dot{\phi}\frac{\partial G_5}{\partial X}-\frac{\partial G_5}{\partial \phi}\right)\right],\\
&{\cal F}_T=2\left[G_4-X\left(\ddot{\phi}\frac{\partial G_5}{\partial X}+\frac{\partial G_5}{\partial \phi}\right)\right]
\end{aligned}
\ee
from (\ref{eq26}) follows that the conditions to avoid ghost and gradient instabilities under tensor perturbations reduce to ${\cal G}_T>0, {\cal F}_T>0$. 
Replacing the solutions (\ref{eq15}) in the expressions for  ${\cal G}_T$ and ${\cal F}_T$ and taking into account the correspondence (\ref{eq25}) we find (using $\kappa^2=M_p^{-2}$)
\be\label{eq29}
{\cal G}_T=M_p^2+\frac{M_p^2}{p(p+1)}\left(\frac{\phi_0^2 t_1^2-8p^2(5p+3)\xi_1}{6M_p^2 t^2}-2p\right)
\ee
\be\label{eq30}
{\cal F}_T=M_p^2-\frac{M_p^2}{p(p+1)}\left(\frac{\phi_0^2t_1^2+8p(p^2-3p-6)\xi_1}{6M_p^2t^2}+2\right)
\ee
where we have used the restrictions (\ref{eq18}). As we are interested in the late time accelerated expansion, we will consider values $p>1$. Then we can qualitatively analyze the conditions for stability under the above discussed approximations at late times for quintessence expansion, i.e. $t=t_1$, $\phi_0\sim M_p$ and $\xi_1\sim-(2p-1)\xi_0/p$. In this case we have
\be\label{eq31}
M_p^2\left(\frac{10p^3+7p^2-9p+1}{6p(p+1)}\right)>0,\,\,\, M_p^2\left(\frac{3p^4+7p^3-7p^2-11p+3}{3p^2(p+1)^2}\right)>0
\ee
which are easily satisfied for $p>1$ (more exactly $p>1.27$). The behavior of ${\cal G}_T$ and ${\cal F}_T$ at asymptotic future time ($t\rightarrow \infty$), as follows from (\ref{eq29}) and (\ref{eq30}) is of the form ${\cal G}_T=\rightarrow M_p^2(p-1)/(p+1)$ and ${\cal F}_T=\rightarrow M_p^2(p^2+p-2)/(p(p+1))$, which are positive provided $p>1$. On the other hand, at asymptotic future time in the de Sitter limit ($p\rightarrow \infty$) we find ${\cal G}_T={\cal F}_T=M_p^2$, so that the de Sitter limit is free of ghost and gradient instabilities under tensor perturbations.\\
\noindent {\bf \it Scalar perturbations}\\
The quadratic action for scalar perturbations $\zeta$ is given by \cite{kobayashi}, \cite{felice1}
\be\label{eq32}
\delta^2S_S=\int dt d^3x a^3\left({\cal G}_S\dot{\zeta}^2-\frac{{\cal F}_S}{a^2}(\vec{\nabla}\zeta)^2\right)
\ee
where
\be\label{eq33}
\begin{aligned}
&{\cal G}_S=\frac{\Sigma}{\Theta^2}{\cal G}_T^2+3{\cal G}_T\\
&{\cal F}_S=\frac{d}{dt}\left(\frac{{\cal G}_T^2}{\Theta}\right)+H\frac{{\cal G}_T^2}{\Theta}-{\cal F}_T
\end{aligned}
\ee
with $\Sigma$ and $\Theta$ given by $(4.25)$, $(4.26)$ in \cite{kobayashi}
For the solution (\ref{eq15}) including the restrictions (\ref{eq17a}) and (\ref{eq18}) the corresponding expressions for 
 ${\cal G}_S$ and ${\cal F}_S$ are too long to be displayed here, and therefore we limit ourselves to the the approximation at late times when the accelerated expansion takes place and $p>1$ ($t\sim t_1$ and $t\rightarrow\infty$) and under the conditions discussed before ($\phi_0\sim M_p$, $\xi_1\sim -(2p-1)\xi_0/p$). The conditions for stability in this case reduce to
\be\label{eq34}
{\cal G}_S=\frac{36 p^9+258 p^8+ 660 p^7 + 575 p^6- 393 p^5- 639 p^4 + 280 p^3+ 42 p^2- 21 p+2}{9 p^2 (1 + p (-6 - 9 p - 2 p^2 + 2 (1 + p)^3))^2}M_p^2>0,
\ee
\be\label{eq35}
{\cal F}_S=\frac{36 p^6+162p^5+132 p^4- 161 p^3- 110 p^2+92 p-11}{36 p^2 (1 + p) (-1 + p (1 + p))^2}M_p^2>0
\ee 
These conditions are satisfied for $0.032< p < 0.5$, $0.564<p<1.2$ and $p>1.33$. Taking the limit $t\rightarrow\infty$ in (\ref{eq33}) one finds
\be\label{eq36}
{\cal G}_S\rightarrow\left(\frac{3(p-1)}{(p+1)(p-2)^2}\right)M_p^2,\,\,\,\, {\cal F}_S\rightarrow\left(\frac{3(p-1)}{p(p^2-p-2)}\right)M_p^2
\ee
So in the far future, in order to avoid ghost and gradient instabilities under scalar perturbations, the power $p$ should satisfy $p>2$, which is consistent with an accelerated universe. We can take the de Sitter limit ($p\rightarrow \infty$) in (\ref{eq34}), (\ref{eq35}) corresponding to the current epoch and also in (\ref{eq36}) at far future. It follows that ${\cal G}_S\rightarrow 0$ and ${\cal F}_S\rightarrow 0$ in both cases, but nevertheless de sound speed is finite since $c_S^2={\cal F}_S/{\cal G}_S\rightarrow 1$. This shows the absence of ghost and gradient instabilities under scalar perturbations in the de Sitter limit.
\section{Discussion}
The power-law solutions in the FRW background represent asymptotic or intermediate states among all possible cosmological evolutions, that might successfully explain different phases 
of the universe evolution. In this paper we have considered power-law solution for the model described by the action (\ref{eq1}). It was shown that in this model the only potential consistent with the power-law evolution is of the Higgs type, where the potential in his broken-symmetry configuration (i.e. with minimum located at $\phi_{min}\ne 0$) describes the matter dominance epoch ($1/2<p<1$), in the limit between decelerated and accelerated expansion ($p=1$) the potential behaves as $V\propto \phi^4$, recovering the symmetric shape, and in the phase of accelerated expansion ($p>1$) the potential continues symmetric. In the de Sitter asymptotic phase the potential behaves as $V\propto\phi^2$.
The symmetric shape of the potential takes place also for the phantom power-law evolution. 
From Eqs. (\ref{eq15}) and (\ref{eq19}) by using the restrictions (\ref{eq17a}) and (\ref{eq18}), we can qualitatively appreciate the contribution of each term in the action (\ref{eq1}) at late times (more precisely at current epoch) in the case when we consider large values of the power $p$ (in fact a value of $p=100/3$ gives the EoS $w=-0.98$) and taking $H_0=p/t_1$, $\phi_0\sim M_p$. Namely, the free kinetic term behaves as $\propto H_0^2M_p^2/p^2$, the kinetic coupling term as $\propto H_0^2M_p^2/p$ and the GB coupling contributes $\propto H_0^2M_p^2/p$ while the potential term (\ref{eq19}) behaves as $V\sim 3H_0^2M_p^2$ (here we used the Eq. (\ref{eq18}) for $\zeta$ and we have assumed $\xi_1\sim -(2p-1)\xi_0/p$). So in this approximation the potential term becomes dominant provided $p>1$. Is worth mentioning that under the approximations made in section 4 leading to conditions (\ref{eq31}), the broken-symmetry shape of the potential becomes unstable since the interval of stability excludes powers $p<1$. Nevertheless this approximation was done for an specific choice of $\xi_1$ and for late time behavior where we assume that the power-law of the type $p>1$ takes place.\\
The present model is a particular case of generalized Galileon theories \cite{deffayet}, \cite{kobayashi}, and we have applied to the model (\ref{eq1}), the perturbative approach developed in \cite{kobayashi}, \cite{felice1}, \cite{felice}, \cite{felice2} for generalized Galileon theories, in order to study the conditions to avoid ghost and gradient instabilities under metric perturbations. We have found the regions of $p$-parameter that satisfy the requirement of stability under metric perturbations, and illustrated the case of quintessence solution at late times. It was also shown that the de Sitter limit is free of ghost and gradient instabilities.\\
Note that the qualitative analysis done in section 3 illustrates the desired scenario, where the current amplitude and mass of the scalar field, are not in conflict with the constraints imposed by the observational data. One might consider for instance the inverse of the parameter $t_1$ as some mass parameter $M$, and change the range of the amplitude and mass of the scalar field which could be relevant in a different cosmological scenario. Of course, the energy scale of the Higgs-type potential considered here is so low with respect to the energy scale of the SM Higgs potential, that there is not measurable connection between the scalar singlet of the present model and the SM Higgs boson. We may think that there is some analogy between both phenomena: the symmetry breaking that gives mass to the particles of the SM, and the symmetry breaking in the considered here potential that delimits the epoch of dominance of matter, where the decelerated expansion takes place. It may be that it is worth studying more thoroughly this analogy and also to implement a mechanism to explain the phase transition in the potential.
\section*{Acknowledgments}
This work was supported by Universidad del Valle under project CI 7890.

\end{document}